\documentclass{article}

\usepackage{arxiv}

\usepackage[utf8]{inputenc} 
\usepackage[T1]{fontenc}    
\usepackage{hyperref}       
\usepackage{url}            
\usepackage{booktabs}       
\usepackage{amsfonts}       
\usepackage{nicefrac}       
\usepackage{microtype}      
\usepackage{lipsum}
\usepackage{graphicx}

\usepackage{caption}
\usepackage{subcaption}
\usepackage{eurosym}
\usepackage{balance}
\usepackage{enumitem}
\usepackage{xcolor}
\usepackage{booktabs} 
\usepackage{cite} 
\newlist{questions}{enumerate}{2}
\setlist[questions,1]{label=RQ\arabic*.,ref=RQ\arabic*}
\setlist[questions,2]{label=(\alph*),ref=\thequestionsi(\alph*)}
\newcommand\res[2]{\scriptsize{$#1$}\tiny{{$\pm #2$}}}

\title{Social Honeypot for Humans: \\Luring People through Self-managed Instagram Pages}

\author{
  Sara Bardi\\
  † University of Padua \\
  Padua, Italy \\
  \texttt{sara.bardi@studenti.unipd.it} \\
   \And
  Mauro Conti \\
  † University of Padua \\
  Padua, Italy \\
  \texttt{conti@math.unipd.it} \\
  \And
  Lucaa Pajola \\
  † University of Padua \\
  Padua, Italy \\
  \texttt{pajola@math.unipd.it} \\
  \And
  Pier Paolo Tricomi\thanks{Contacting author.} †\\
  Chisito S.r.l. \\
  Padua, Italy \\
  \texttt{tricomi@math.unipd.it} \\
}

\begin{document}
\maketitle
\begin{abstract}
Social Honeypots are tools deployed in Online Social Networks (OSN) to attract malevolent activities performed by spammers and bots. To this end, their content is designed to be of maximum interest to malicious users.
However, by choosing an appropriate content topic, this attractive mechanism could be extended to \textit{any} OSN users, rather than only luring malicious actors. 
As a result, honeypots can be used to attract individuals interested in a wide range of topics, from sports and hobbies to more sensitive subjects like political views and conspiracies.
With all these individuals gathered in one place, honeypot owners can conduct many analyses, from social to marketing studies. 
\par
In this work, we introduce a novel concept of social honeypot for attracting OSN users interested in a generic target topic. 
We propose a framework based on fully-automated content generation strategies and engagement plans to mimic legit Instagram pages.
To validate our framework, we created 21 self-managed social honeypots (i.e., pages) on Instagram, covering three topics, four content generation strategies, and three engaging plans. In nine weeks, our honeypots gathered a total of 753 followers, 5387 comments, and 15739 likes.
These results demonstrate the validity of our approach, and through statistical analysis, we examine the characteristics of effective social honeypots.

\end{abstract}


\section{Introduction}

In recent years, Social Network Analysis (SNA) has emerged as a powerful tool for studying society. The large amount of relational data produced by Online Social Networks (OSN) has greatly accelerated studies in many fields, including modern sociology~\cite{smith2020social}, biology~\cite{fisher2017social}, communication studies~\cite{hagen2018crisis}, and political science~\cite{kim2021planetary}.  
SNA success can be attributed to the exponential growth and popularity OSN faced~\cite{Alexa}, with major OSN like Facebook and Instagram (IG) having billions of users~\cite{karl,statista}. Researchers developed a variety of tools for SNA~\cite{rani2021survey}; however, elaborating the quintillion bytes of data generated every day~\cite{dailydata} is far from trivial~\cite{brooker2016have}. The computational limitations compel scientists to conduct studies on sub-samples of the population, often introducing bias and reducing the quality of the results~\cite{boyd2012critical}. 
Furthermore, the reliability of data is hindered by adversarial activities perpetuated over OSN~\cite{jain2021online, conti2022captcha}, such as the creation of fake profiles~\cite{sheikhi2020efficient}, crowdturfing campaigns~\cite{wang2012serf, tricomi2022we}, or spamming \cite{hu2014online,zhu2012discovering,murugan2018detecting}. 

Back in the years, cybersecurity researchers proposed an innovative approach to overcome the computational limitation in finding malicious activity in OSN (e.g., spamming), by proposing social honeypots~\cite{webb2008social,lee2010uncovering,stringhini2010detecting}: profiles or pages created ad-hoc to lure adversarial users, analyze their characteristics and behavior, and develop appropriate countermeasures. Thus, their search paradigm in OSN shifted from ``look for a needle in the haystack'' (i.e., searching for spammers among billions of legit users) to ``the finer the bait, the shorter the wait'' (i.e., let spammers come to you). 

\paragraph{Motivation}

The high results achieved by such techniques inspired us to generalize the approach, gathering in a \textit{single} place \textit{any target users} we wish to study.
Such a framework's uses are various, from the academic to the industrial world. 
First, \textit{profilation} or \textit{marketing} toward target topics: IG itself provides page owners to know aggregated statistics (e.g., demographic) of their followers and users that generate engagement.\footnote{Instagram API provides to the owner aggregated statistics of followers (gender, age, countries) when their page reaches 100 followers~\cite{IG_API}.}
Second, \textit{social cybersecurity analytics}: researchers or police might deploy social honeypots on sensitive themes to attract and analyze the behavior of people who engage with them. 
Examples of themes are fake news and extremism (e.g., terrorism). 
Although our ``general'' social honeypot may be used either benignly (e.g., to find misinformers) or maliciously (e.g., to find vulnerable people to scam), in this paper, we only aim to examine the feasibility of such a tool, and its effectiveness.
Moreover, we investigate whether this technique can be fully automated, limiting the significant effort of creating a popular IG page~\cite{robertson2018instagram}. We focus on IG given its broad audience and popularity. Furthermore, IG is the most used social network for marketing purposes, with nearly 70 percent of brands using IG influencers (even virtual~\cite{conti2022virtual}) for their marketing campaigns~\cite{infl_m}.


\paragraph{Contribution}
In this work, we present an automated framework to attract and collect legitimate people in social honeypots. 
To this aim, we developed several strategies to understand and propose guidelines for building effective social honeypots. Such strategies consider both \textit{how to generate content automatically} (from simple to advanced techniques), and \textit{how to engage with the OSN} (from naive to complex interactions).
In detail, we deployed 21 honeypots and maintained them for nine weeks. Our four content generation strategies involve state-of-the-art Deep Learning techniques, and we actively engage with the network following three engagement plans.  

The main contributions of our paper can be summarized as follows:
\begin{itemize}
    \item We define a novel concept of Social Honeypot, i.e., a flexible tool to gather \textit{real people} on IG interested in a target topic, in contrast to previous studies focusing on malicious users or bots; 
    \item We propose four automatic content generation strategies and three engagement plans to build self-maintained IG pages;
    \item We demonstrate the quality of our proposal by analyzing our 21 IG social honeypots after a nine weeks period. 
\end{itemize}

\paragraph{Outline}
We begin our work discussing related works (§\ref{Background}). Then, we present our methodology and implementation in §\ref{sec:methodology} and §\ref{sec:impl}. In §\ref{sec:eval}, we evaluate the effectiveness of our honeypots, while §\ref{sec:social} presents social analyses. We discuss the use cases of our approach and its challenges in §\ref{sec.disc} and conclude the paper in §\ref{sec:concl}.

\section{Related Works}
\label{Background}

\label{honey}
\subsubsection{Honeypot} Honeypots are decoy systems that are designed to lure potential attackers away from critical systems~\cite{stallings2012computer}. Keeping attackers in the honeypot long enough allows to collect information about their activities and respond appropriately to the attack. Since legit users have no valid reason to interact with honeypots, any attempt to communicate with them will probably be an attack. Server-side honeypots are mainly implemented to understand network and web attacks~\cite{john2011heat}, to collect malware and malicious requests~\cite{yegneswaran2005architecture}, or to build network intrusion detection systems~\cite{kreibich2004honeycomb}. Conversely, client-side honeypots serve primarily as a detection tool for compromised (web) servers~\cite{moshchuk2006crawler, wang2006automated}. 

\subsubsection{Social Honeypot} Today, honeypots are not limited to fare against network attacks. Social honeypots aim to lure users or bots involved in illegal or malicious activities perpetuated on Online Social Networks (OSN). Most of the literature focused on detecting spamming activity, i.e., unsolicited messages sent for purposes such as advertising, phishing, or sharing undesired content~\cite{stringhini2010detecting}. The first social honeypot was deployed by Webb et al.~\cite{webb2008social} on MySpace. They developed multiple identical honeypots operated in several geographical areas to characterize spammers' behavior, defining five categories of spammers. 
Such work was extended to Twitter by Lee et al. in 2010~\cite{lee2010uncovering}, identifying five more spammers' categories, and proposing an automatic tool to distinguish between spammers and legit users.   
Stringhini et al.~\cite{stringhini2010detecting} proposed a similar work on Facebook, using fake profiles as social honeypots. Similarly to previous works, these profiles were passive, i.e., they just accepted incoming friend requests. Their analysis showed that most spam bots follow identifiable patterns, and only a few of them act stealthily. 
De Cristofaro et al.~\cite{de2014paying} investigated Facebook Like Farms using social honeypots, i.e., blank Facebook pages. In their work, they leveraged demographic, temporal, and social characteristics of likers to distinguish between genuine and fake engagement. 
The first ``active'' social honeypot was developed on Twitter by Lee et al.~\cite{lee2011seven}, tempting, profiling, and filtering content polluters in social media. These social honeypots were designed to not interfere with legitimate users' activities, and learned patterns to discriminate polluters and legit profiles effectively. 60 honeypots online for seven months gathered 36'000 interactions.
More active social honeypots were designed by Yang et al.~\cite{yang2014taste}), to provide guidelines for building effective social honeypots for spammers. 96 honeypots online for five months attracted 1512 accounts. Last, pseudo-honeypots were proposed by Zhang et al.~\cite{zhang2019toward}, which leveraged already popular Twitter users to attract spammers efficiently. They run 1000 honeypots for three weeks, reaching approximately 54'000 spammers.


\subsubsection{Differences with previous work} To date, social honeypots have been mainly adopted to detect spammers or bot activities. The majority of research focused on Twitter, and only a few works used other social networks like Facebook. There are several reasons behind this trend. First, spamming is one of the most widespread malicious activities on social networks because it can lead to other more dangerous activities.
Second, Twitter APIs and policies facilitate data collection, and there are widely adopted Twitter datasets that can be used for further analysis. To the best of our knowledge, there are no works that utilize social honeypots on Instagram, perhaps because it is difficult to distribute, maintain and record honeypots' activities on this social network. Moreover, our goal is to attract \textit{legit users} rather than spammers, which is radically different from what was done insofar. Indeed, many analyses could be easier to conduct by gathering people in one place (e.g., an IG page). For instance, a honeypot could deal with peculiar topics to simplify community detection~\cite{bedi2016community}, could advertise a product to grasp consumer reactions~\cite{campbell2014segmenting}, understand political views~\cite{mcclurg2003social}, analyze and contrast misinformation~\cite{del2016spreading}, conspiracies~\cite{ahmed2020covid}, and in general, carry out any Social Network Analytics task~\cite{dey2018social}. Last, owners of IG pages can see the demographic information of their followers (inaccessible otherwise), having extremely helpful (or dangerous) information for further social or marketing analyses~\cite{singh2019analysis}.


\section{Methodology}
\label{sec:methodology}
\subsection{Overview \& Motivation}
\begin{figure*}[ht!]
    \centering
    \includegraphics[width=.95\linewidth]{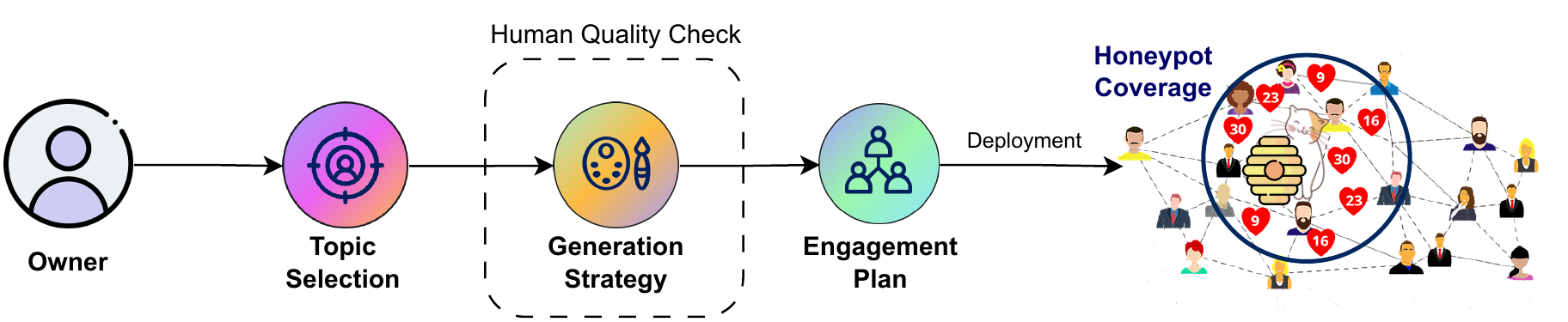}
    \caption{Pipeline overview to create a social honeypot. After the owner decides on the topic, generation strategy, and engagement plan, the honeypot automatically generates posts to interact with the social network. After the post is automatically generated, the owner can approve it or request a new one to meet the desired quality.}
    \label{fig:SH-overview}
\end{figure*}

The purpose of our social honeypots is to attract people interested in a target topic. 
The methodology described in this section is intended for Instagram (IG) pages, but it can be extended to any generic social network (e.g., Facebook) with minor adjustments. 
We define the social honeypot as a combination of three distinct components: 
(i) the honeypot \textit{topic} that defines the theme of the IG page (§\ref{ssec:/methodology/topic-selection});  
(ii) the \textit{generation strategy} for creating posts related to a target topic (§\ref{ssec:/methodology/generation-strategy}); 
(iii) the \textit{engagement plan} that describes how the honeypot will engage the rest of the social network (§\ref{subsec:engplans}).
Figure~\ref{fig:SH-overview} depicts the social honeypot pipeline. 

\par
Our study examines different types of honeypots with a variety of topics, generation strategies, and engagement plans, outlined in the rest of this section.
Our experiments aim to answer the following research questions:

\begin{questions}
    \item Can self-managed social honeypots generate engagement on Instagram?
    \item How do the topic selection, post generation strategy, and engagement plan affect the success of a social honeypot?
    \item How much effort (computation and costs) is required to build an effective social honeypot?
\end{questions}
The remainder of the section describes the strategies we adopt in our investigation, along with technical implementation details.

\subsection{Topic Selection}\label{ssec:/methodology/topic-selection}
Building a honeypot begins with selecting the topic of its posts. Such a choice will impact the type of users we will attract. 
The topic's nature might vary, from hobbies and passions like sports and music to sensitive issues like political views and conspiracies.  
As an example, if we wish to promote a new product of a particular brand, the topic might be the type of product we intend to promote. Alternatively, if we intend to develop a tool for spam detection, we should choose a topic that is interesting to spammers. This will ensure that they will be attracted to the honeypot's content.
We can even design honeypots with generic topics that can be used for marketing profiling or social studies. In conclusion, the topic should be chosen in accordance with the honeypot's ultimate purpose. 

\subsection{Post Generation Strategies}\label{ssec:/methodology/generation-strategy}
The generative process aims to create posts pertaining to the honeypot topic. 
A two-part artifact is produced: the \textit{visual} component of the post (i.e., the image), and its \textit{caption}. 
We propose four distinct methods to generate posts, each with its own characteristics and algorithms. For ethical reasons, we excluded techniques that might violate the author's copyright (e.g., re-posting). However, unscrupulous honeypot creators could conveniently use these strategies. In this section, we provide the strategies high-level view to serve as a framework. For technical implementation details (e.g., the actual models we used), please refer to Appendix~\ref{appendix:impl}. Since this stage involves deep generative models that might produce artifacts affecting the post quality, the owner can approve a post or request a new one with negligible effort. 


\begin{figure*}[t!]
\centering
\begin{subfigure}[b]{0.45\textwidth}
    \centering
    \includegraphics[width=.99\linewidth]{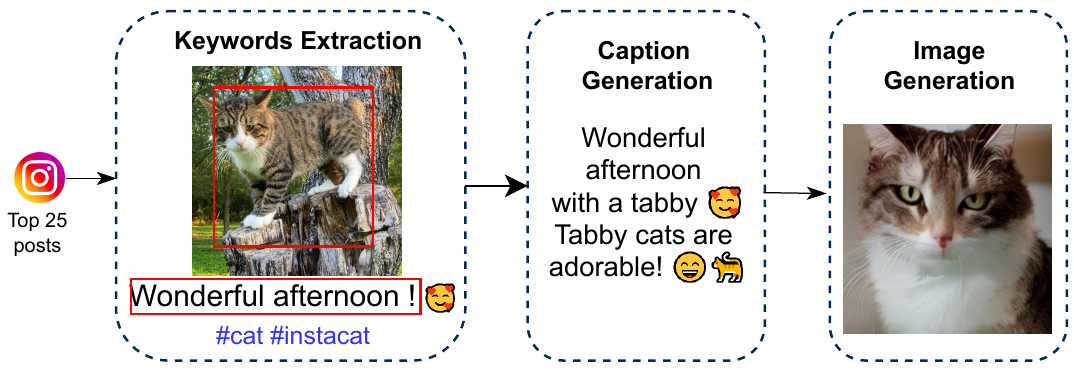}  
    \caption{InstaModel}
    \label{fig:insta-model}
\end{subfigure}
\begin{subfigure}[b]{0.45\textwidth}
    \centering
    \includegraphics[width=.90\linewidth]{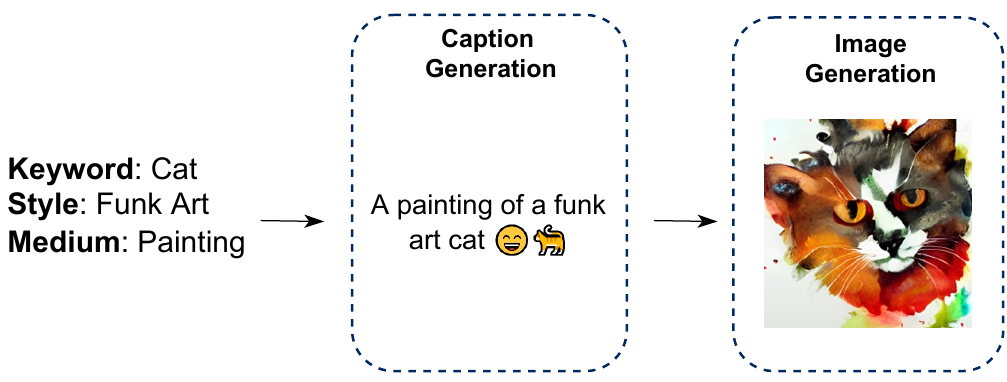} 
    \caption{ArtModel}
    \label{fig:art}
\end{subfigure}
\begin{subfigure}[b]{0.453\textwidth}
    \centering
    \includegraphics[width=.99\linewidth]{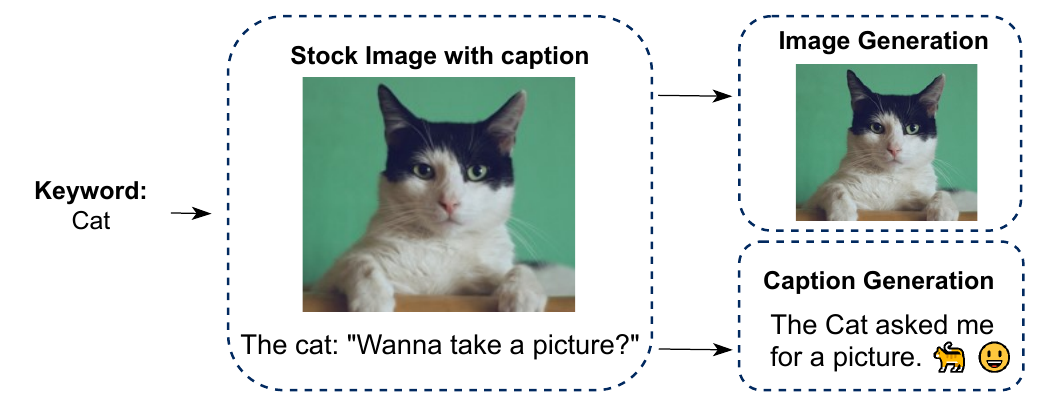}  
    \caption{UnsplashModel}
    \label{fig:u}
\end{subfigure}
\begin{subfigure}[b]{0.45\textwidth}
    \centering
    \includegraphics[width=.99\linewidth]{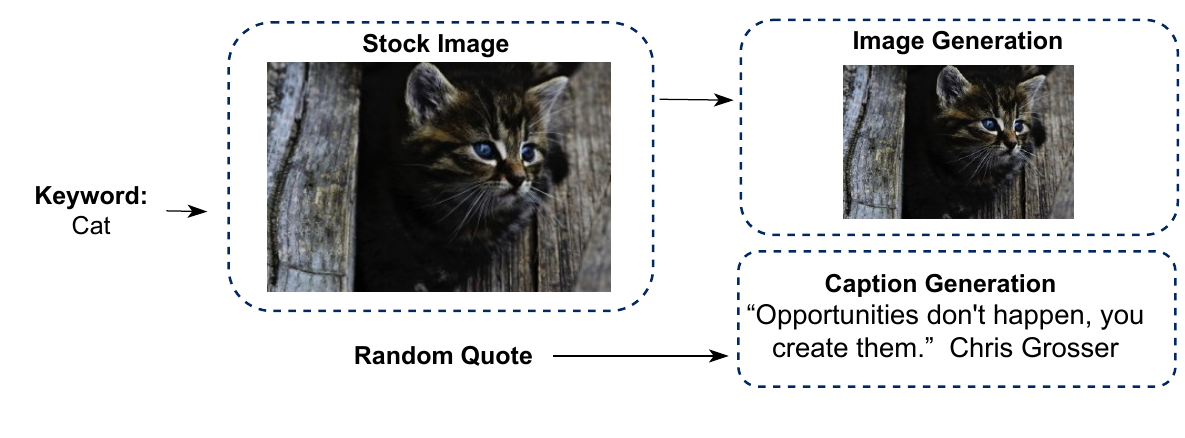}  
    \caption{QuotesModel}
    \label{fig:q}
\end{subfigure}
\caption{Overview of Post Generation strategies.}
\label{FIGURE LABEL}
\end{figure*}

\subsubsection{InstaModel}
\textit{InstaModel} is a generative schema that leverages machine learning techniques to generate both images and captions. Figure~\ref{fig:insta-model} shows its overview.
The schema begins by retrieving one starting post among the 25 most popular IG posts for a popular hashtag related to the honeypot topic.\footnote{Starting from the main topic hashtags (i.e., \#cat, \#food, \#car), we daily create the set of hashtags contained in the top 25 posts, from which we draw the hashtag to retrieve the starting post.} Next, the pipeline performs, in order, caption generation and image generation steps. 
\begin{itemize}
    \item \textit{Caption Generation.}
    The algorithm uses an \textit{Object Detector} tool\footnote{Object detectors are Computer Vision-based tools that identify objects composing a given scene. Each object is accompanied by a probability score.} to extract the relevant elements of the starting post's image. In the absence of meaningful information (e.g., is a meme or unrelated to the topic)\footnote{We discard those images that do not contain at least a topic-related element with a high probability. }, we discard that image. When this occurs, the algorithm restarts and uses another sample from the top 25. 
    If the image is kept, the algorithm uses the list of resulting elements (i.e., keywords) to generate a sentence, leveraging a \textit{keyword-to-text} algorithm. Note that we discard from the keywords list those elements with very low probability. 
    The output of the \textit{keyword-to-text} phase (i.e., the new caption) is further refined to align with IG captions, for example, by adding emojis and hashtags, as presented in §\ref{subsec:engplans}. 
    \item \textit{Image Generation.} The caption generated in the previous step serves as input to produce the post image. To achieve this goal, we use \textit{text-to-image} models, i.e., algorithms that produce more images from a single input. An operator would choose the most appropriate option or a random option in such a case. 
    We remark that \textit{InstaModel} severely adopts generative models. 
    Indeed, we used state-of-the-art computer vision, NLP, and image generation models for object detection, text generation, and image generation, respectively. 
\end{itemize}

\subsubsection{ArtModel}
\textit{ArtModel} leverages the ability of novel \textit{text-to-image} generative models (e.g., DALL-E) to interpret artistic keywords as inputs. Figure~\ref{fig:art} shows the overview of the model. Similarly to InstaModel, the process starts by generating a caption, and, subsequently, the image.
\begin{itemize}
    \item \textit{Caption Generation.} Differently from \textit{InstaModel}, the input to generate the caption does not come from other IG posts. Instead, we randomly select the target keyword (e.g., cat), the artistic style of the picture (e.g., Picasso, impressionism), and a medium (e.g., painting, sketch). We create a single sentence by filling pre-defined templates with such three keywords, and add emojis and hashtags as for \textit{InstaModel}.
    
\item \textit{Image Generation.} Similar to \textit{InstaModel}, the caption (without emojis and hashtags) serves as input for a \textit{text-to-image} model, which generates the final image.  

\end{itemize}
\par

\subsubsection{UnsplashModel}
This algorithm employs DL models only to generate the caption.
In opposition to \textit{InstaModel} and \textit{ArtModel}, \textit{UnsplashModel} starts from the image generation, and then generates the caption (Figure~\ref{fig:u}).

\begin{itemize}
    
\item \textit{Image Generation.} The image is randomly selected by a stock images website -- in this case, Unsplash\footnote{https://unsplash.com/}. The search is based on a randomly selected keyword that reflects the target topic, from a list defined by the owner. 

\item \textit{Caption Generation.} Unsplash images are usually accompanied by captions free of license. We further refine the caption with a \textit{rephrase} model, and add emojis and hashtags as for the previous models.
\end{itemize}


\subsubsection{QuotesModel}
Last, we present \textit{QuotesModel}, a variant of \textit{UnsplashModel}, presented in Figure~\ref{fig:q}.
The objective of this strategy is to determine whether AI-based techniques are necessary to generate attractive IG posts. Therefore, this model does not involve the use of artificial intelligence to create captions and images. In addition, using quotes to caption photos is a diffused strategy~\cite{quotes_cap}. 

\begin{itemize}
    
\item \textit{Image Generation.} The image generation process is the same as \textit{UnsplashModel}, involving stock images.

\item \textit{Caption Generation.} Captions are randomly selected by popular quotes from famous people (e.g., `Stay hungry, stay foolish' -- Steve Jobs). Quotes are retrieved from a pool with 1665 quotes~\cite{quotes}.

\end{itemize}

\subsection{Engagement Plans}\label{subsec:engplans}
Lastly, the engagement plan defines how the social honeypot interacts with the rest of the social network (e.g., other users or pages). 
We defined three plans, varying in effort required to maintain interactions, and whether paid strategies are involved: 
\begin{itemize}
    \item \textit{PLAN 0}: low interactions and no paid strategies;
    \item \textit{PLAN 1}: high interactions and no paid strategies;  
    \item \textit{PLAN 2}: high interactions and paid strategies. 

\end{itemize}



\subsubsection{PLAN 0}
The plan does not involve automatic interactions with the rest of the social network. At most, the owner replies to comments left under the honeypot's posts. The plan uses the well-known \textit{Call To Actions} (CTA)~\cite{cta} in the posts. Such a strategy consists in creating captions that stimulate users' engagement (e.g., liking, commenting, sharing the post). Examples are captions containing simple questions (e.g., `How was your day?'), polls and quizzes (e.g., `What should I post next?), or exhorting users to share their opinions (e.g., `What do you think about it?'). Following the caption best strategies for IG posts~\cite{hashtags}, 
we added 15 random hashtags related to our topic, 8 with broad coverage and 7 with medium-low coverage. More details about the hashtags selections in Appendix~\ref{appendix:impl}.
In this plan, paid strategies are not involved.


\subsubsection{PLAN 1}
The plan is a variant of \textit{PLAN 0} with explicit social networking interactions. We call these actions \textit{spamming}. 
The spamming consists of automatically leaving likes and comments on the top 25 posts related to the topic (as described in \textit{InstaModel}). Comments resemble legit users (e.g., `So pretty!') and not spammers (e.g., `Follow my page!'), and were randomly picked from a list we manually created by observing comments usually left under popular posts.
The goal of such activities is to generate engagement with the owner of popular posts, hoping to redirect this stream to the honeypot. 
When a user follows us, we follow back with a probability of 0.5, increasing the page's number of followings, resembling a legit page.  
During our experiments, we also adopted a more aggressive (and effective) spamming strategy called \textit{Follow \& Unfollow} (F\&U)~\cite{foll_unf}, consisting in randomly following users, often causing a follow back, and then remove the following after a couple of days. To not be labeled as spammers, we constantly respected the balance \# following $<$ \#followers.
In this plan, paid strategies are not involved.


\subsubsection{PLAN 2}
This plan increments \textit{PLAN 1} with two paid strategies.

\paragraph{Buying followers} When we create a honeypot, we buy $N$ followers. In theory, highly followed pages might encourage users to engage more, and gain visibility from IG algorithm~\cite{instaalg2022}. Therefore, we aim to understand if an initial boost of followers can advantage honeypots. Such followers will be discarded during our analyses. We set $N = 100$, and we buy passive followers only.\footnote{Passive followers only follow the page, but they do not engage further.}

\paragraph{Content sponsoring} IG allows posts' sponsoring for a certain amount of time. The target population can be automatically defined by IG, or chosen by the owner w.r.t. age, location, and interests. Since we are interested in studying the population attracted by our content, rather than attracting a specific category of users, we let IG decide our audience, directly exploiting its algorithms to make our honeypots successful.

\section{Implementation}
\label{sec:impl}
\subsection{Topic Selection}
We investigate the honeypots' effectiveness over three distinct topics: \textit{food}, \textit{cat}, and \textit{car}. 
We selected such topics to account for different audience sizes, measured by coverage levels. 
Coverage is a metric that counts the total number of posts per hashtag or, in other words, the total number of posts that contain that hashtag in their captions. This information is available on IG by just browsing the hashtag.
More in detail, we selected: \textbf{Food} (high coverage, \#food counts 493 million posts),
\textbf{Cat} (medium coverage, \#cat counts 270 million posts), and \textbf{Car} (low coverage, \#car counts 93 million posts).
We chose these topics, and not more sensitive ones, mainly for ethical reasons. Indeed, we did not want to boost phenomena like misinformation or conspiracies through our posts, nor identify people involved in these themes. However, we designed our methodology to be as general as possible, and adaptable to any topic with little effort.



\subsection{Testbed}
\label{sec:experiment}
We deployed 21 honeypots on Instagram, seven for each selected topic (i.e., food, cat, and car), that we maintained for a total of nine weeks.
Within each topic, we adopt all post generation strategies and engagement plans.
For the post generation strategies, three honeypots use both InstaModel and ArtModel, three honeypots use UnsplashModel and QuotesModel, and one honeypot combines the four. 
Such division is based on the image generation strategy, i.e., if images are generated with or without Deep Learning algorithms. All posts were manually checked before uploading them on Instagram to prevent the diffusion of harmful or low-quality content. This was especially necessary for AI-generated content, whose low quality might have invalidated a fair comparison with non-AI content.\footnote{The effort for the honeypot manager is limited to a quick approval, which could not be necessary with more advanced state-of-the-art models, e.g., DALL-E 2~\cite{dall-e2} or ChatGPT~\cite{chatGPT}.}
Similarly, for the engagement plan, two honeypots adopt PLAN 0, two PLAN 1, and three PLAN 2. 
Table~\ref{tab:honeypot} summarizes the 21 honeypots settings. Given the nature of our post generation strategies and engagement plans, we set as baselines the honeypots involving \textit{UnsplashModel + QuotesModel} as generation strategy and \textit{PLAN 0} as engagement plan (h1, h8, h15). Indeed, these honeypots are the simplest ones, requiring almost no effort from the owner. Setting baselines is useful to appreciate the results of more complex methods, given that there are currently no baselines in the literature.
\par
By following the most common guidelines ~\cite{2postPerDay,2postPerDay2}, each honeypot was designed to publish two posts per day, with at least 8 hours apart from each other.
\par
During the nine weeks of experiments, we varied PLAN 1 and PLAN 2. In particular, we started PLAN 1 with spamming only, and PLAN 2 with buying followers.
During the last week, both plans adopted more aggressive strategies, specifically, PLAN 1 applied F\&U techniques, while PLAN 2 sponsored the two most-popular honeypot posts for one week, paying \EUR{2}/day for each post.
For our analyses, we collected the following information:
\begin{itemize}
    \item Total number of followers per day;
    \item Total number of likes per post;
    \item Total number of comments per post.
\end{itemize}
Moreover, IG API provided the gender, age, and geographical locations of the audience when applicable, as explained in §\ref{ssec.audience}.

\begin{table}[!ht]
\centering
\scriptsize
\caption{Honeypots deployed.}
\begin{tabular}{@{}ccc@{}}
\toprule
\textbf{ID}         & \multicolumn{1}{c}{\textbf{Post Generation Strategy}} & \textbf{Engagement Plan} \\ \toprule
 \multicolumn{3}{c}{\textit{food}} \\
 \midrule
h1 (baseline)     & UnsplashModel + QuotesModel & PLAN 0 \\
h2            & UnsplashModel + QuotesModel                           & PLAN 1                   \\
h3     & UnsplashModel + QuotesModel & PLAN 2 \\
h4  & InstaModel + ArtModel       & PLAN 0 \\
h5      & InstaModel + ArtModel       & PLAN 1 \\
h6    & InstaModel + ArtModel       & PLAN 2 \\
h7  & All Models                  & PLAN 2 \\

 \midrule
 \multicolumn{3}{c}{\textit{cat}}\\ \midrule
h8 (baseline)     & UnsplashModel + QuotesModel & PLAN 0 \\
h9        & UnsplashModel + QuotesModel & PLAN 1 \\
h10       & UnsplashModel + QuotesModel & PLAN 2 \\
h11      & InstaModel + ArtModel       & PLAN 0 \\
h12   & InstaModel + ArtModel       & PLAN 1 \\
h13       & InstaModel + ArtModel       & PLAN 2 \\
h14       & All Models                  & PLAN 2 \\  \midrule
\multicolumn{3}{c}{\textit{car}} \\ \midrule
h15  (baseline)   & UnsplashModel + QuotesModel & PLAN 0 \\
h16       & UnsplashModel + QuotesModel & PLAN 1 \\
h17     & UnsplashModel + QuotesModel & PLAN 2 \\
h18   & InstaModel + ArtModel       & PLAN 0 \\
h19      & InstaModel + ArtModel       & PLAN 1 \\
h20          & InstaModel + ArtModel       & PLAN 2 \\

h21     & All Models                  & PLAN 2 \\ \bottomrule
\end{tabular}
\label{tab:honeypot}
\end{table}

\paragraph{Implementation Models}
In §\ref{sec:methodology} we presented a general framework to create social honeypots. In our implementations, we employed deep learning state-of-the-art models in several steps. To extract keywords in  \textit{InstaModel} we adopted InceptionV3~\cite{szegedy2016rethinking} as object detector, pre-trained on ImageNet~\cite{deng2009imagenet} with 1000 classes. From the original caption, we extracted nouns and adjectives through NLTK python library\footnote{\url{https://www.nltk.org/}}. As \textit{keyword-to-text} algorithm, we adopted Keytotext~\cite{keytotext} based on T5 model~\cite{raffel2020exploring}; while for \textit{text-to-image} processes we opted for Dall-E Mini~\cite{Dayma_DALL}. Finally, in \textit{UnsplashModel}, the rephrase task was performed using the Pegasus model~\cite{zhang2019pegasus}.
\section{Honeypots Evaluation}
\label{sec:eval}



\subsection{Overall Performance}\label{subsec:ovper}
The first research question \textit{RQ1} is whether social honeypots are capable of generating engagement. 
After nine weeks of execution, our 21 social honeypots gained: 753 followers (avg 35.86 per honeypot), 5387 comments (avg 2.01 per post), and 15730 likes (avg 5.94 per post). 
More in detail, Table~\ref{tab:honeypot-results} (left side) shows the overall engagement performance at the varying of our three variables, i.e., topic, generation strategy, and engagement plan. 
The reader might notice that not only our honeypots \textit{can} generate engagement, answering positively to the \textit{RQ1}, but that also topic, generation strategy, and engagement plan have different impacts to the outcomes. 
For instance, \textit{cat} honeypots tend to have higher followers and likes, while \textit{car} ones generate more comments. 
Similarly, \textit{non-AI} generation methods tend to have higher likes, as well as \textit{PLAN 1}. 
We investigate the effect of different combinations later in this section. 

\begin{table}[!ht]
\centering
\scriptsize
\caption{Honeypots overall performance. On the left side, we report the average (and std) engagement generated by the honeypots. On the right, we report the number of honeypots with a non-stationary trend. The results are reported based on the topic, generation strategy, and engagement plan.}
\vspace{10pt}
\begin{tabular}{@{}c|ccc|ccc@{}}
\toprule
& \multicolumn{3}{c}{\textit{\textbf{Average Engagement}}} & \multicolumn{3}{c}{\textit{\textbf{Engagement Trend}}} \\
\toprule
& \textbf{\#Followers}        & \multicolumn{1}{c}{\textbf{\#Comments}} & \textbf{\#Likes} 
& \textbf{\#Followers}         & \multicolumn{1}{c}{\textbf{\#Comments}} & \textbf{\#Likes} \\ \toprule
 \multicolumn{7}{c}{\textit{topic}} \\
 \midrule
food & \res{38.5}{33.7} & \res{216.4}{18.5} & \res{698.4}{139.7} & 6/7 & 3/7 & 7/7\\
cat  & \res{\textbf{47.4}}{17.5} & \res{182.1}{23.5}  & \res{\textbf{923.1}}{214.8} & 6/7 & 2/7 & 4/7\\
car  & \res{21.9}{9.7} & \res{\textbf{371.0}}{26.2} & \res{625.6}{96.6} & 7/7 & 3/7 & 6/7\\
 \midrule
 \multicolumn{7}{c}{\textit{generation strategy}}\\ \midrule
AI      & \res{37.9}{30.9} & \res{248.4}{94.6} & \res{654.2}{138.3} & 7/9 & 4/9 & 6/9 \\
non-AI  & \res{32.7}{21.3} & \res{\textbf{264.2}}{90.6} & \res{\textbf{842.5}}{235.2} & 9/9 & 3/9 & 8/9\\ 
Mixed   & \res{\textbf{39.3}}{7.9} & \res{257.7}{80.0} & \res{753.0}{125.9} & 3/3 & 1/3 & 3/3\\\midrule
\multicolumn{7}{c}{\textit{engagement plan}} \\ \midrule
PLAN 0 & \res{11.5}{8.4} & \res{\textbf{266.0}}{105.8} & \res{641.3}{210.7} & 4/6 & 4/6 & 5/6\\
PLAN 1 & \res{\textbf{60.0}}{25.2} & \res{254.2}{94.3} & \res{\textbf{835.2}}{210.7} & 6/6 & 2/6 & 4/6 \\
PLAN 2 & \res{36.0}{14.0} & \res{251.8}{79.1} & \res{763.4}{206.1} & 9/9 & 2/9 & 8/9\\ \bottomrule
\end{tabular}
\label{tab:honeypot-results}
\end{table}

\subsection{Honeypot Trends Analysis}\label{sub:trends}
Social honeypots can generate engagement, but we are further interested in understanding trends of such performance:  \textit{is honeypots' engagement growing over time?}
A honeypot with a positive trend will likely result in a higher future attraction. 
On the opposite, a stationary trend implies limited opportunities to improve. 
\par
\begin{figure*}[!htpb]
    \centering
    \includegraphics[width = .95\linewidth]{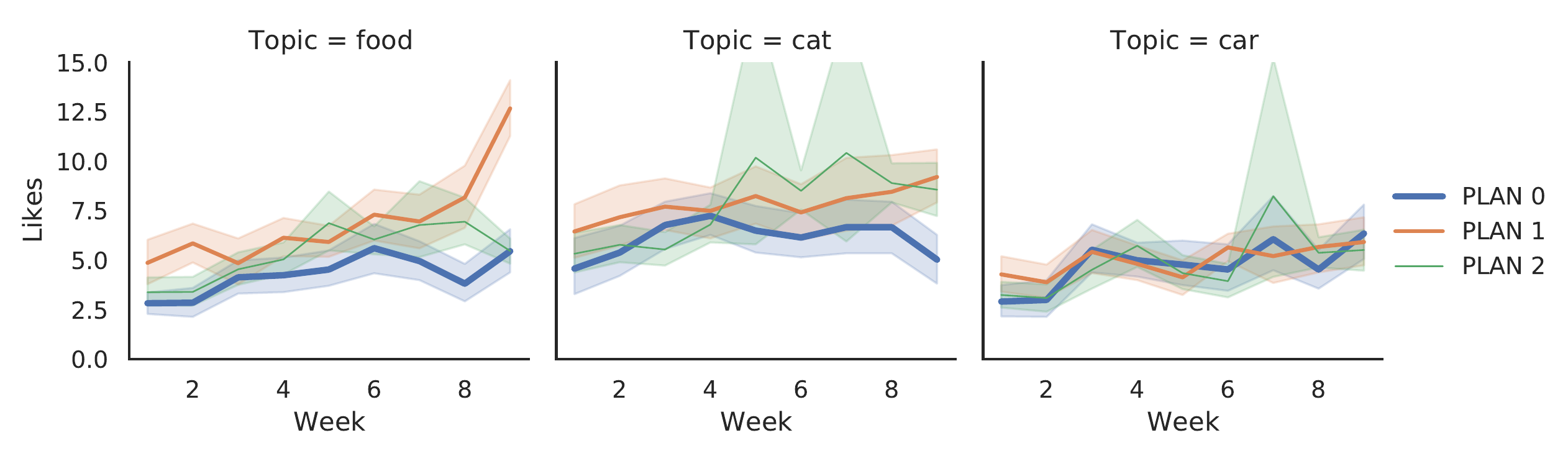}
    \caption{Likes trend of our honeypots grouped by engagement plan.}
    \vspace{-2em}
    \label{fig:food-trend}
\end{figure*}

The qualitative analysis reported in Figure~\ref{fig:food-trend} motivates the trend investigation.
The figure presents the average number of Likes per post gained by our honeypots over time, grouped by engagement plan. In general, PLAN 1 honeypots tend to attract more likes as they grow, followed by PLAN 2 and PLAN 0, in order. In particular, a constantly increasing number of likes is shown by honeypots with PLAN 1, especially for food-related pages: starting from an average of $\sim$5 likes per post (week 1st) to $\sim$12.5 likes per post (week 9th).  
We evaluate the presence of stationary trends by adopting the \textit{Augmented Dickey–Fuller test} (ADF)~\cite{mushtaq2011augmented}. 
In this statistical test, the null hypothesis $H_0$ suggests, if rejected, the presence of a non-stationary time series. On the opposite, the alternative hypothesis $H_1$ suggests, if rejected, the presence of a stationary time series. 
We conducted the statistical test for each honeypot and the three engagement metrics: \#Followers, \#Likes, and \#Comments. 
A $p$-value $>$ 0.05 is used as a threshold to understand if we fail to reject $H_0$.
Table~\ref{tab:honeypot-results} (right side) reports the result of the analysis. 
The number of Followers and Likes is non-stationary in 19 and 17 cases out of 21, respectively. 
Conversely, the number of comments per post is stationary in most of the honeypots. 
This outcome suggests that engagement in terms of likes and followers varies over time (positively or negatively), while the number of comments is generally constant. As shown in Figure~\ref{fig:food-trend}, and given the final number of followers higher than 0 (i.e., at creation time), we can conclude that our honeypots present, in general, a growing engagement trend.

\subsection{The Impact of Honeypots Configuration}\label{subsec:config}
We now investigate whether the three variables (i.e., topic, generation strategy, and engagement plan) have a statistical impact on the success of the honeypots, answering \textit{RQ2} and \textit{RQ3}. Given the stationary trend of comments, we focus solely on likes per post and followers per honeypot.

\subsubsection{Likes}
\begin{figure*}[!htpb]
    \centering
    \includegraphics[width = .95\linewidth]{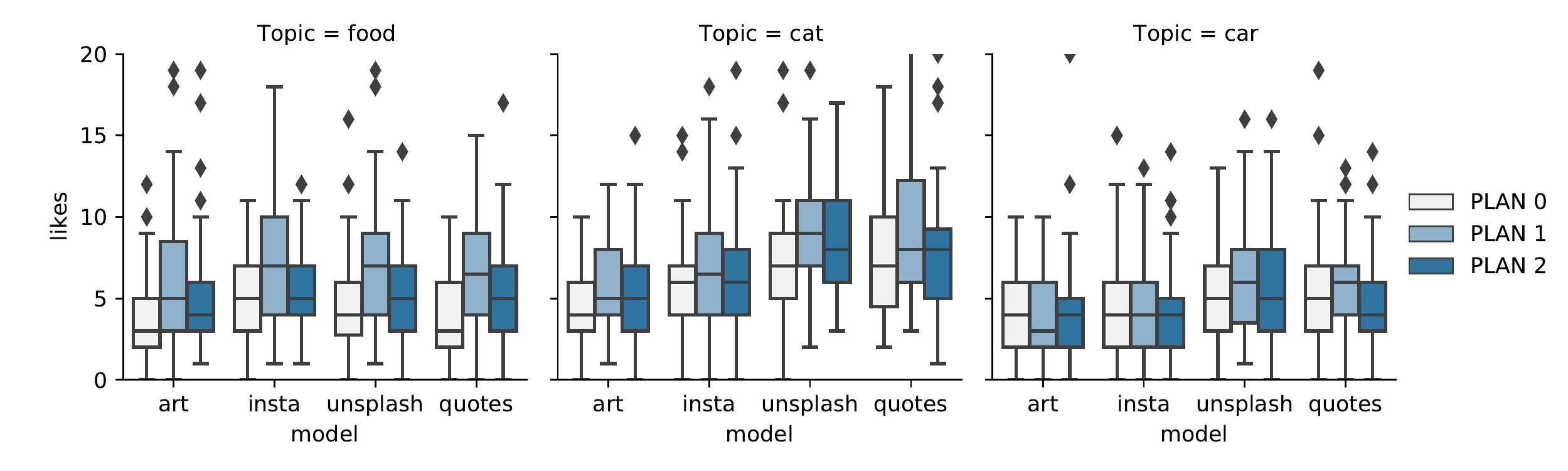}
    \caption{Distribution of likes at the varying of topic, model generation strategy, and engagement plan.}
\vspace{-2em}
    \label{fig:three-factors}
\end{figure*}

Figure~\ref{fig:three-factors} depicts the distribution of honeypots Likes at the varying of the topic, generation strategy, and engagement plan.   
In general, there is a difference when the three variables are combined. 
For example, on average, honeypots belonging to cats, with non-AI generative models, and with PLAN1 or PLAN2 have higher values than the rest of the honeypots. 
Moreover, in general, honeypots adopting PLAN1 have higher results. 
\par
To better understand the different impacts the three variables have on Likes, we conducted a three-way ANOVA. We found that both topic, engagement plan, and generation strategy are significantly ($p$-{value} $<$ 0.001) influencing the Likes. Furthermore, we found significance even in the combination of topic and engagement plan ($p$-{value} $<$ 0.001), but not in the other combinations. This result confirms the qualitative outcomes we have presented so far.
We conclude the analysis by understanding which topic, generation strategy, and engagement plan are more effective. To this aim, we performed Tukey's HSD (honestly significant difference) test with significance level $\alpha = 5\%$. 
Among the three topics, \textit{cat} is significantly more influential than both \textit{food} and \textit{car} ($p$-value = 0.001). 
Regarding the generation strategies, non-AI-based models (i.e., UnsplashModel and InstaModel) outperform AI-based ones. 
Last, PLAN1 and PLAN2 outperform PLAN0 ($p$-value = 0.001), while the two plans do not show statistical differences between them.

\subsubsection{Followers}
Tukey's HSD test revealed statistical differences in the number of followers as well. 
For the analysis, we use the number of followers of each honeypot at the end of the 9th week. 
We found that \textit{cat} statistically differ from \textit{car} ($p$-value $<$ 0.01), while there are no significant differences between \textit{cat} and \textit{food}, or \textit{food} and \textit{car}. 
Regarding the generation strategy, we found no statistical difference among the groups. 
Finally, all three engagement plans have a significant impact on the number of followers ($p$-value~=~0.001), where PLAN 1 $>$ PLAN 2 $>$ PLAN 0. 

\subsubsection{Aggressive engagement plans}
We recall that honeypots deployed with PLAN 1 and PLAN 2 adopted more aggressive engagement strategies on week 9th: \textit{Follow \& Unfollow} for PLAN 1, and \textit{Content Sponsoring} for PLAN 2.
Thus, we investigated whether aggressive plans result in more engagement in terms of comments, likes, and followers.
The analysis is performed with Tukey's HSD (honestly significant difference) test with significance level $\alpha = 5\%$. 
We found no statistical difference in comments in PLAN 1 and PLAN 2.
On the opposite, the average number of likes per post shows a statistically significant improvement in PLAN1 ($p$-value = 0.01): on average, 7.44 and 9.17 likes per post in weeks 8th and 9th, respectively. 
No statistical difference is found for PLAN 2; indeed, only the sponsored content benefited (i.e., a few posts).\footnote{All sponsored content belongs to weeks before the 9th.}
Last, we analyze the difference between the total amount of followers at the end of weeks 8th and 9th. 
PLAN 1 honeypots \#Followers moved, on average, from ${45.7}\pm{19.1}$ of week 8th, to ${60.7}\pm{26.2}$ of week 9th, with no statistical difference. 
PLAN 2 honeypots \#Followers moved, on average, from ${22.3}\pm{11.6}$ of week 8th, to ${30.7}\pm{13.9}$ of week 9th. The difference is statistically supported ($p$-value $<$ 0.05).

\subsection{Baseline Comparison}\label{subsec:baseline}
Social honeypots are effective, depending on topics, generation strategies, and engagement plans. Since we are the first, to the best of our knowledge, to examine how to attract \textit{people} using social honeypots (not bots or spammers), there are no state-of-the-art baselines to compare with. 
Therefore, we compare our methodology with (i)
our proposed non-AI generative models with a PLAN 0 engagement strategy (baseline) and (ii) real Instagram pages trends.  

\subsubsection{Baseline}
\begin{figure}[!htpb]
    \centering
    \includegraphics[width = .95 \linewidth]{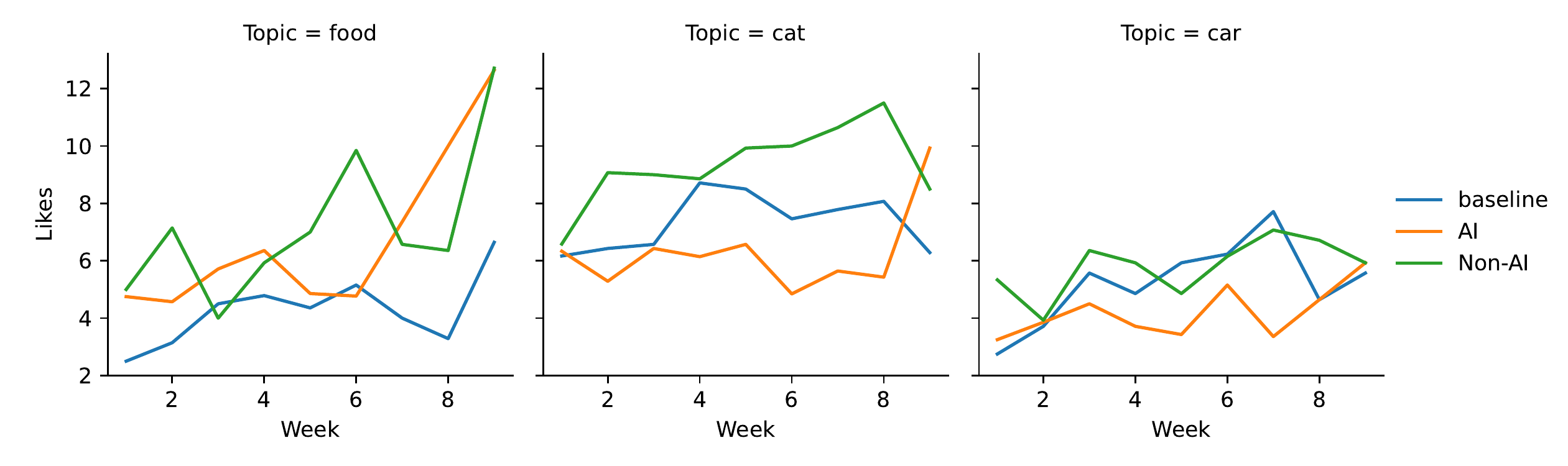}
    \caption{Baseline comparison (average likes) with PLAN1 social honeypots.}
    \label{fig:baseline-comp}
\end{figure}

\begin{figure}[!htpb]
    \centering
    \includegraphics[width = .95 \linewidth]{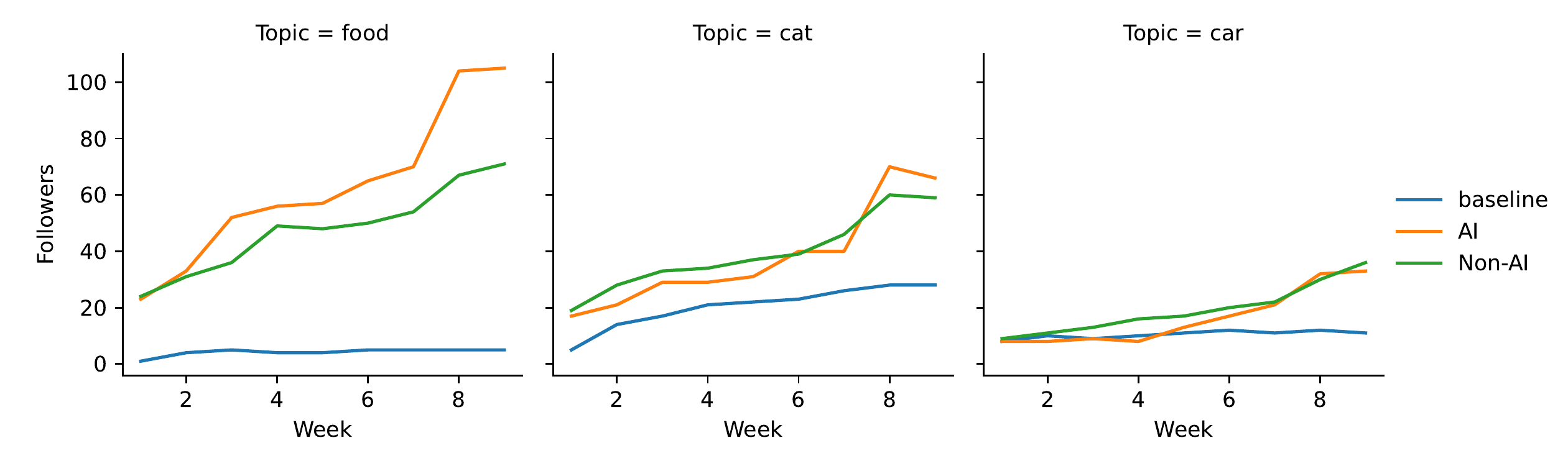}
    \caption{Baseline comparison (followers) with PLAN1 social honeypots.}
\vspace{-2em}
    \label{fig:baseline-comp-followers}
\end{figure}

This represents the most simplistic method someone might adopt: adding stock images, with random quotes, without caring about the engagement with the rest of the social network. 
From §\ref{subsec:config}, we statistically showed that the definition of engaging plans is essential to boost engagement in social honeypots. 
We remark on this concept with Figures~\ref{fig:baseline-comp} and~\ref{fig:baseline-comp-followers} that show the comparison among the baselines and PLAN 1 social honeypot -- which are the most effective ones -- in terms of likes and followers over the 9 weeks: in terms of AI and Non-AI strategies, our advanced honeypots outperform in 3 out of 6 cases and 6 out of 6 cases the baselines for likes and followers, respectively. 
Such results confirm the remarkable performance of our proposed framework. 
Our strategies might perform worse than the baselines (regarding likes) when the image quality is unsatisfactory. Indeed, as demonstrated in our prior work~\cite{tricomi2023follow}, likes on IG are usually an immediate positive reaction to the post's image. Since Unsplash images are usually high-quality and attractive, they might have been more appealing than AI-generated images in these cases. 
\par
Although comparing our approach with other social honeypots~\cite{lee2011seven,yang2014taste,zhang2019toward} carries some inherent bias (the purpose and social networks are completely different), we still find our approach aligned with (or even superior than) the literature. Lee et al.~\cite{lee2011seven} gained in seven months through 60 honeypots a total of $\sim$36000 interactions (e.g., follow, retweet, likes), which is approximately 21.5 interactions per honeypot/week. Our honeypots reached a total of 21870 interactions, which is approximately 115.7 interactions per honeypot/week, i.e., more than five times higher. Yang et al.~\cite{yang2014taste} lured 1512 accounts in five months using 96 honeypots, i.e., 0.788 accounts per honeypot/week. We collected 753 followers, which is 3.98 accounts per honeypot/week, i.e., five times higher. Last, Zhang et al.~\cite{zhang2019toward} carefully selected and harnessed 1000 popular Twitter accounts (which they called pseudo-honeypots) for three weeks to analyze spammers. Giving these accounts were already heavily integrated into the social network, they reached over 476000 users, which is around 159 accounts per (pseudo-)honeypot per week. We remind that the purpose of these comparisons is to give an idea of the effectiveness of other social honeypots rather than to provide meaningful conclusions.    


\vspace{-2em}

\subsubsection{Instagram Pages}
We now compare our PLAN 1 social honeypots with real IG public accounts. 
Accordingly, we analyzed the first nine weeks of activities on popular IG pages related to food, cat, and cars.
We selected nine popular IG pages for each topic, 3 with $\sim10K$ followers, 3 with $\sim100K$ followers, and 3 with more than a million followers. 
We collected the number of comments and likes for each post published during this period. 
Due to IG limitations, we could access only information at the time of collection, implying that posts might be a few years old. 
Monitoring new pages would be meaningless since we do not know a priori whether they will become popular. 
\par
We noticed that it is impossible to compare such baselines with our social honeypots because, generally, the considered IG pages contain posts with hundreds of likes and comments even in their first week of activity. For instance, $+1M$ pages' first posts reached more than 2000 likes.
Possible explanations behind this phenomenon are: 
(i) the considered 18 pages were already popular before their creation (e.g., on a different or older OSN like Facebook); 
(ii) the considered 18 pages massively sponsored all their content; 
(iii) we are facing the \textit{earlybird bias}, where older posts contain not just engagement from the first nine weeks, but also engagement from later periods, even years.\footnote{Earlybird bias appears in other social contexts like online reviews~\cite{liu2007low}.} 
To further explain this phenomenon, we contacted such IG pages (we extended our survey to 36 pages). Questions focused on the first weeks of activity.\footnote{For instance, we asked whether the page resulted from an already existing page (on IG or other platforms),  or the strategies they adopted to manage the pages (e.g., spam, sponsoring).}
Unfortunately, up to the submission date, none of the contacted pages replied. 
\par
Although there is no evidence in the literature on how long it takes to make an Instagram page famous, most sources consider the initial growth (from 0 to 1000 followers) to be the most challenging part~\cite{hardpart1,hardpart2}, with an overall monthly growth rate of about 2\%~\cite{IGgrowth}. Furthermore, success requires lots of dedication to follow best practices consistently~\cite{practices}, which is extraordinarily time-consuming and far from trivial. Being in line with these trends in a fully automated and effortless manner is already an impressive achievement. Our work can serve as a baseline and inspiration for future work.

\section{Social Analyses}
\label{sec:social}
\subsection{Comments analysis}
An interesting (and unexpected) result is that, without the premeditated intention of building spammer detectors, most of the comments we received came from spammers. To estimate the total number of spam comments, we first manually identified patterns used by spammers on our honeypots (e.g., expressions like ``send pic'' or ``DM us''). Afterward, using a pattern-matching approach, we found that 95.33\% of the comments we received on our social honeypots came indeed from spammers. All spammers' accounts shared similar behavior in commenting: (i) there was always a mention `@' to other accounts, and (ii) they commented almost immediately after the post creation. Such considerations suggest these accounts are bots that target many recent posts, perhaps searching by specific hashtags. Such findings indicate that fresh pages could be a powerful tool to detect spammers with \textit{minimal} effort. We also highlight that spam comments are a well-known issue that affects the majority of IG pages~\cite{spam} and is not limited to our honeypot pages. Therefore, we argue that creating pages that do not attract spammers is nearly impossible. Nevertheless, IG itself is employing and improving automatic screening mechanisms~\cite{ig_detectors, ig_authenth} to limit such behavior. When such mechanisms are enhanced, our honeypots will become more accurate.

\subsection{Followers analysis}
As most of our comments were spam, we investigated whether followers were the same.
We manually inspected the followers of our most followed social honeypot for each topic, identifying three categories of followers:
\begin{itemize}
    \item \textit{Real people:} users that publish general-topic posts, with less than 1000 followers\footnote{After 1000 followers, users are considered nano influencers~\cite{tiers}.}, and real profile pictures; 
    \item \textit{Pages and Influencers:} users that publish topic-specific posts (e.g., our honeypots) or with more than 1000 followers;
    \item \textit{Bots:} users whose characteristics resemble a bot, following well-known guidelines~\cite{akyon2019instagram}, e.g., fake or absent profile picture, random username, highly imbalanced follower/following count, zero or few ($<$ 5) posts.
\end{itemize}
From Table~\ref{tab:followerAnalysis}, we notice the three honeypots have different audiences. The \textit{food} honeypot obtained the most real followers, \textit{car} reached more bots, and \textit{cat}, was followed mainly by pages. These results confirmed that (i) our honeypots can reach real people, (ii) the audience category depends on the topic, and (iii) spammers' threat is limited to comments. On an interesting note, most pages following our \textit{cat} honeypot were cat-related pages.
\vspace{-2em}
\begin{table}[!h]
\centering
\footnotesize
\caption{Percentage of real people, pages, and bots for the best social honeypot in each topic. }
\begin{tabular}{@{}lccc@{}}
\toprule
 & \multicolumn{1}{l}{Real People} & \multicolumn{1}{l}{Pages} & \multicolumn{1}{l}{Bots} \\ \midrule
Food & \textbf{48,08}\% & 37,50\% & 14,42\% \\
Cat  & 10,61\% & \textbf{72,72}\% & 16,67\% \\
Car  & 30,30\% & 21,21\% & \textbf{48,49}\% \\ \bottomrule
\end{tabular}%
\vspace{-2em}
\label{tab:followerAnalysis}
\end{table}

\subsection{Reached Audience}\label{ssec.audience}
We conclude the experimental results with a detailed analysis of the audience our honeypots reached. 
In particular, we performed two distinct analyses: (i) \textit{Honeypot reached audience},  and (ii) \textit{Sponsored posts audience}, i.e., IG features available for honeypots with 100 followers and sponsored content, respectively. 
After nine weeks of computation, one honeypot satisfies the requirement of 100 followers (honeypot ID: h9). About the sponsored content, we obtained information about 9 posts (one per honeypot belonging to PLAN~2).

\subsubsection{Honeypot audience} The honeypot h9 (topic: food, generation strategy: AI, and engagement plan: PLAN 1) gained 103 followers: the majority is distributed over the age range $[25 - 34]$ with 32\% (equally distributed among men and women),  $[35, 44]$ with 10\% of women and 27\% of men. Most followers came from India (11.7\%), Bangladesh (10.7\%), and Japan (9.7\%).

\subsubsection{Sponsored posts audience}
For this analysis, we recall that we set our sponsoring strategy leveraging the automatic algorithm provided by IG. Overall, sponsored posts achieved great success in terms of generated engagement. 
On average, food posts reached 30.6, 116, and 60.6 likes for food, cat, and car posts, respectively. These numbers are strongly above the average likes per post 5.9. 
IG offers an analytic tool to inspect the reached audience; this feature perfectly fits in the scope of social honeypots, since it allows finding insights about the attracted audience. 
For each post, the following information is available: quantitative information (i.e., reached people, likes, comments, sharing, saved), and demographic distribution in percentage (gender, age, location).
The detailed report is available in Appendix~\ref{app:anal}. 
We observed interesting trends: 
\begin{itemize}
    \item \textit{food} audience: the gender is almost balanced (female audience slightly more attracted), and the predominant age range is 18-34. Top locations: Campania, Lombardia, and Puglia.\footnote{IG automatic algorithm maximized the audience toward authors country, i.e., Italy, reporting Italian regions.}
    \item \textit{cat} audience: the gender distribution is toward the female sex, and the predominant age range is 18-34. Top locations: Emilia Romagna, Lombardia, Piemonte. 
    \item \textit{car} audience: the gender is strongly distributed toward the male sex, and the predominant age range is 18-24. Top locations: Lazio, Lombardia. 
\end{itemize}
To conclude, with minimal effort (i.e, \EUR{2}/day per post), an owner can get useful information, e.g., to use in marketing strategies..
\section{Toward a Real Implementation} \label{sec.disc}
So far, we have demonstrated our social honeypots can attract real people in a fully automated way. With little effort, they can already be deployed for an array of situations.  In this section, we first reason about the use cases of our approach, highlighting both positive and negative outcomes. Then, we present the current challenges and limitations of implementing this work in real scenarios. 

\subsection{Use Cases} 
Our work aims to show the lights and shadows of social networks such as Instagram. 
People can easily deploy automated social honeypots that can attract engagement from hundreds or even thousands of users. 
Upon on that, analyses on these (unaware) users can be conducted. 
As cyber security practitioners, we know that this technology might be exploited not only for benign purposes, but also to harm users~\cite{9699411}. 
Therefore, this work contributes to the discussion about the responsible use of online social networks, in an era when technologies like artificial intelligence are transforming cyber security. 
We list in this section possible social honeypot applications. 


\subsubsection{Marketing}
The first natural adoption of our proposed social honeypots is for marketing purposes. 
Suppose someone is interested in understanding ``who is the average person that loves a specific theme'', where themes might be music, art, puppies, or food. 
With a deployed social honeypot, the owner can then analyze the reached audience by using the tools offered by IG itself (as we ethically did in this paper) or by further gathering (potentially private) information on the users' profile pages~\cite{10.1145/3508398.3511517}.  

\subsubsection{Phishing and Scam}
Similarly to marketing, social honeypots can be used by adversaries to conduct phishing and scam campaigns on IG users. 
For instance, the social honeypot might focus on cryptocurrency trading: once identified potential victims, attackers can target them aiming to obtain sensitive information (e.g., credentials), or to lure them into fraudulent activities such as investment scams, rug pulls, Ponzi schemes, or phishing.  

\subsubsection{Spammer Identification}
Social honeypots can also be created to imitate social network users, by posting content and interacting with other users. As we noticed in our experiments, they can attract spammers. Therefore, our proposed framework can be adopted by researchers to spot and study new types of spamming activities in social networks. 

\subsubsection{Monitoring of Sensible Themes}
An interesting application of social honeypots is to identify users related to sensible themes and monitor their activities (within the honeypot). 
Examples of such themes are fake news and extremism~\cite{10.1145/3522756}. 
Researchers or authorities might leverage social honeypots to identify users that actively follow and participate in such themes, and then carefully examine their activity. 
For instance, honeypot owners can monitor how people respond to specific news or interact inside the honeypot. 

\subsection{Challenges and Limitations}
The first challenge we faced in our work is the massive presence of spammers on IG. Most of them are automated accounts that react to particular hashtags and comments under a post for advertisement or scamming purposes~\cite{zhang2017instagram, IGSpam}. 
This factor can inevitably limit our approach when we aim to gather only real people. As a countermeasure, honeypots should include a spam detector (e.g.~\cite{zhang2017instagram, haqimi2019detection}) to automatically remove spammers. 
On the contrary, this approach could be useful directly to reduce the spamming phenomenon. Many pages can be created with the purpose of attracting spammers and reporting them to IG for removal.

The second challenge we encountered is the lack of similar works in the literature. Because of this, we have no existing baselines to compare with, and it could be difficult to understand whether our approach is truly successful. However, in nine weeks, we obtained more than 15k likes and gathered $\sim750$ followers in total, which is not trivial as discussed in §\ref{subsec:baseline}. Our most complex methods surpassed the simplest strategies we identify, which can serve as a baseline and source of inspiration for future works.

Among the limitations, we inspected only generic (and ethical) topics. A comprehensive study in this direction would give much more value to our work, especially dealing with delicate topics (e.g., conspiracies, fake news). Moreover, our approach is currently deployable on IG, but would be hard to transfer to other platforms. Even if this can be perceived as a limitation, it would be naive to consider all social media to be the same. Indeed, each of them has its own content, purpose, and audience. Developing social honeypots for multiple platforms can be extremely challenging, which is a good focus for future research. Last, there was no clear connection between the posts of our honeypots. When dealing with specific topics, it might be necessary to integrate more cohesive content.

\section{Conclusions}
\label{sec:concl}

The primary goal of this work was to first understand the feasibility of deploying self-managed Instagram Social Honeypots, and we demonstrated that \textit{it is possible} in §\ref{subsec:ovper}. Moreover, from the results obtained in our analyses we can derive the following outcomes and guidelines:
\begin{enumerate}
    \item \textit{Topics} plays an important role in the success of the honeypot. 
    \item \textit{Generation strategies} does not require complex DL-based models, but simple solutions such as stock images are enough. Similarly, we saw that posts containing random quotes as captions are as effective as captions describing the content; 
    \item \textit{Engagement plan} is essential. We demonstrated that a naive engagement strategy (PLAN 0) results in a low volume of likes and followers. Moreover, the engagement plan without costly operations (PLAN 1) works as well as plans involving followers acquisition and content sponsoring; 
    \item \textit{Sponsored content} is a useful resource to preliminary assess the audience related to a specific topic; 
    \item Social honeypots not only attract \textit{legitimate} users, but also \textit{spammers}. As a result, they can be adopted even for cybersecurity purposes. Future implementation of social honeypots might include automatic tools to distinguish engagement generated by legitimate and illegitimate users.  
\end{enumerate}
In conclusion, we believe that our work can represent an important milestone for future researchers to easily deploy and collect social network users' preferences. New research directions might include not only general topics like cats and food, but more sensitive themes like fake news, or hate speech. In the future, we expect generative models to be always more efficient (e.g., DALL-E 2~\cite{dall-e2} or ChatGPT~\cite{chatGPT}), thus increasing the reliability of our approach (or perhaps making it even more dangerous).


\subsection*{Ethical Considerations}
\label{sec:ethics} 
Our institutions do not require any formal IRB approval to carry out the experiments described herein. Nonetheless, we designed our experiments to harm OSN users as less as possible, adhering to guidelines for building Ethical Social Honeypots~\cite{dittrich2015ethics}, based on the Menlo report~\cite{bailey2012menlo}. Moreover, we dealt with topics (cars, cats, food) that should not hurt any person's sensibility. In our work, we faced two ethical challenges: data collection and the use of deception. 
Similar to previous works~\cite{lee2011seven,yang2014taste,de2014paying}, we collected only openly available data (provided by Instagram), thus no personal information was extracted, and only aggregated statistics were analyzed. Moreover, all information is kept confidential and no-redistributed. Upon completion of this study, all collected data will be deleted. This approach complies with the GDPR. To understand the honeypot's effectiveness, similar to previous works, we could not inform users interacting with them about the study, to limit the Hawthorne effect~\cite{franke1978hawthorne}. However, we will inform the deceived people at the end of the study, as suggested by the Menlo report.

\bibliographystyle{unsrt}  
\bibliography{references}  

\appendix
\clearpage
\section{Implementation details}
\label{appendix:impl}
\subsection{Models}
In this appendix we will describe how InstaModel, ArtModel, UnsplashModel and QuotesModel were implemented. All of them have different characteristics but, at the same time, share some common functionalities that will be explained before of the actual implementation of the four models.

\paragraph{Shared functionalities}
One of the shared functionalities is adding emojis to the generated text. This is done with a python script which scans the generated caption trying to find out if there are words that can be translated with the corresponding emoji. To make this script more effective, it looks also for synonyms of nouns and adjectives found in the text to figure out if any of them can be correlated to a particular emoji. As last operation, the script chooses randomly, from a pool of emojis representing the "joy" sentiment, one emoji for each sentence that will be append at the end of each of them. 

CTA are simple texts that may encourage a user to do actions. These CTA are sampled randomly from a  manually compiled list and then added at the end of the generated caption.

The last shared feature is the selection of hashtags. As said before, through the Instagram Graph API we are able to get the first 25 posts for a specific hashtag and from them we extracted all the hashtags contained in the caption. Thus we compiled an hashtag list for each of the three topic sorted from the most used to the least used. Instagram allows to insert at most 30 hashtags in each posts but we think that this number is too high with respect to the normal user's behavior. For this reason, we decided to choose 15 hashtags that are chosen with this criteria: 8 hashtags are sampled randomly from the first half of the list in the csv file, giving more weight to the top ones, while the other 7 are sampled randomly from the second half of the list, giving more weight to the bottom part of the list. The intuition is that we are selecting the most popular hashtags together with more specific hashtags.  

\paragraph{InstaModel}
Starting from the caption generation, InstaModel uses the Instagram Graph API to retrieve the top 25 posts for a specific hashtag. In practice, the chosen hashtag will be the topic on which the corresponding honeypot is based. Once we have all the 25 posts, they are checked to save only those that have an English caption before being passed to the object detector block. The object detector is implemented by using the InceptionV3 model for object detection tasks.  InceptionV3 detects, in the original image, the object classes with the corresponding accuracy and if the first's class score is not greater than or equal to 0.25, the post will be discarded. Otherwise, the other classes are checked as well and only if their scores are greater than 0.05 will be considered as keywords for the next step. Regarding the original caption, nouns and adjectives are extracted by using nltk python library. Notice that words such as "DM" or "credits" and adjectives such as "double" or similar, are not considered. This is because they usually belong to part of the caption that is not useful for this process. 

Keyword2text\footnote{\url{https://huggingface.co/gagan3012/k2t}} is the NLP model that transforms a list of keywords in a preliminary sentence. This preliminary sentence is then used by OPT model to generate the complete text. Considering the computational resources available to us, the model used is OPT with 1.3 billion parameters. We suggest to save the text generated by OPT in a file text because it will be used subsequently to generate the corresponding image. Once we have the complete generated text, emojis are added together with a CTA sentence that is standard in any post. The last step for caption generation is to append hashtags: they will be chosen by sampling from the corresponding csv file with the reasoning mentioned above. 

The last step of InstaModel is image generation and for this purpose Dall-E Mini (\cite{Dayma_DALL}) is used. The prompt will be the text generated after the OPT stage, the one that has been save separately. It is relevant to highlight that the process with Dall-E Mini is not completely automatic and there should be a person that choose the most suitable image for the giving caption. 

\paragraph{ArtModel}
ArtModel starts from a prompt generated with a python script and uses Dall-E mini, like InstaModel, to generate the corresponding image. The style and the medium are chosen randomly from two lists. Example of styles can be "cyberpunk", "psychedelic", "realistic" or "abstract" while examples of medium are "painting", "drawing", "sketch" or "graffiti". The topic of the honeypot is used as subject of the artistic picture generated by Dall-E Mini. 
Once the image is generated, the prompt, added of emojis, CTA and the corresponding hashtags, will be used as Instagram caption. 

\paragraph{UnsplashModel}
UnslashModel does not generate images but uses stock images retrieved from the Unsplash websites. Unsplash has been chosen not only because it gives the opportunity to find images together with the relative captions, but also because it offers API for developers that can be used easily. To avoid reusing the same images more than once, each image's id is saved in a text file which will be checked at each iteration. 
For the caption generation, the original caption is processed by Pegasus model (\cite{zhang2019pegasus}) which is an NLP model quite good in the rephrase task. As always, emojis, CTA and hashtags are added to the final result.

\paragraph{QuotesModel}
QuotesModel makes use of Pixabay\footnote{\url{https://pixabay.com/}} stock images website to avoid reusing Unsplash even for this model. Also in this case, we use the topic of the specific honeypot as query tag. As for UnsplashModel, to avoid reusing the same image for different posts, once we have downloaded the image, its id is saved in a text file which will be checked every time needed. 
For the caption generation, a quote is sampled randomly from a citation dataset~\cite{quotes_cap}. In this case, the model does not add emojis to the text because we think that the quote, by itself, can be a valid Instagram caption. On the contrary, as always, CTA and hashtags are added to the text.


\begin{table*}[!h]
\centering
\caption{Overview of the sponsored content attracted users}
\resizebox{0.65\linewidth}{!}{
\begin{tabular}{l|lll|lll|lll}
\toprule
\multicolumn{10}{c}{\textit{\textbf{Overview}}}\\\midrule
\textit{honeypot}         & h3  & h6  & h7  & h10 & h13 & h14 & h17 & h20 & h21 \\
\textit{topic}            & food & food  & food  & cat & cat & cat & car & car & car \\
\textit{gen. strat.}& AI & NON AI  & NON AI & AI & NON AI & NON AI & AI & NON AI & NON AI \\
\textit{audience}         & 3126 & 3412 & 5337 & 3245 & 4597 & 2863 & 10698 & 6824 & 9633 \\
\textit{likes}            & 21 & 34 & 37 & 118 & 163 & 67 & 20 & 25 & 127 \\
\textit{comments}         & 1 & 3 & 7  & 3 & 8 & 1 & 3 & 11 & 3 \\
\textit{saved}            & 1 & 0 & 21 & 12 & 29 & 7 & 2 & 6 & 44 \\ \midrule
\multicolumn{10}{c}{\textit{\textbf{Gender Coverage $[\%]$}}} \\\midrule
\textit{women}            & 42.2 & 60.0 & 87.8 & 67.2 & 67.7 & 59.0 & 8.6 & 8.7 & 5.6 \\
\textit{men}              & 57.0 & 38.7 & 11.7 & 31.5 & 30.7 & 39.3 & 89.5 & 90.7 & 93.6 \\ \midrule
\multicolumn{10}{c}{\textit{\textbf{Age Coverage}} $[\%]$} \\\midrule
$13-17$                   & 0.1 & 0.1 & 0 & 0 & 0 & 0.1 & 0.2 & 0.1 & 0.1 \\
$18-24$                   & 39.1 & 37.7 & 35.9 & 20.8 & 33.8 & 38.6 & 64.3 & 45.7 & 52.5 \\
$25-34$                   & 29.8 & 12.9 & 36.0 & 21.2 & 25.2 & 15.2 & 12.7 & 31.8 & 26.8 \\
$35-44$                   & 14.5 & 11.6 & 14.3 & 15.6 & 13.0 & 12.4 & 6.5 & 10.8 & 9.4 \\
$45-54$                   & 9.0 & 18.3 & 8.2 & 18.7 & 14.0 & 13.7 & 8.1 & 5.1 & 6.1 \\
$55-64$                   & 4.7 & 12.9 & 3.8 & 15.8 & 9.3 & 12.4 & 5.0 & 3.6 & 3.0 \\
$65+$                     & 2.5 & 6.0 & 1.3  & 7.5 & 4.3 & 7.2 & 2.9 & 2.6 & 1.8 \\ \midrule
\multicolumn{10}{c}{\textit{\textbf{Geographic Coverage}} $[\%]$} \\\midrule
\textit{Campania}         & 14.7 & 11.3 & 9.1  & N.A.  & N.A. & 8.7 & 7.8 & 8.7 & N.A. \\
\textit{Emilia-Romagna}   & N.A. & N.A & N.A. & 9.7 & 8.7 & 9.2 & N.A. & 8.6 & 9.2 \\
\textit{Lazio}            & N.A. & 7.9 & 8.3  & 9.4 & 10.5 & N.A. & 8.2 & 11.1 & 9.5 \\
\textit{Lombardia}        & 12.4 & 12.0 & 13.2 & 19.6 & 18.8 & 17.2 & 14.0 & 19.0 & 20.9 \\
\textit{Piemonte}         & N.A. & N.A. & N.A. & 9.0 & 8.5 & 7.5 & N.A.  & N.A. &  8.0\\
\textit{Puglia}           & 12.5 & 10.9 & 8.9  & N.A. & N.A. & N.A. & 8.9 & N.A. & N.A. \\
\textit{Sicilia}          & 9.0 & 10.0 & 9.2 & N.A. & N.A. & N.A. & 10.4 & N.A. & N.A. \\
\textit{Tuscany}          & N.A. & N.A. & N.A. & 7.2 & N.A. & N.A. & N.A. & N.A. & N.A. \\
\textit{Veneto}           & 9.0 & N.A. & N.A. & N.A. & 7.7 & 8.4 & N.A. & 8.8 & 10.1 \\
\bottomrule
\end{tabular}
}
\label{tab:sponsoring}
\end{table*}

\subsection{Spamming}
\label{sec:spamming}
Honeypots with PLAN 1 or PLAN 2 engagement plans will automatically interact with the posts of other users. The idea is to retrieve the top 25 Instagram posts for the hashtag corresponding to the specific topic of the honeypot and like and comment each of them.

For the implementation we used Selenium which is a tool to automates browsers and it can be easily installed with pip command. Selenium requires a driver to interface with the chosen browser and in our case, since we chose Firefox, we have downloaded the geckodriver. The implementation consists of a python class which has three main methods: \texttt{login}, \texttt{like\_post} and \texttt{comment\_post}

The login method is invoked when the honeypot accesses to Instagram. 
The like\_post method searches, in the DOM, for the button corresponding to the like action and then it clicks it.
The comment\_post method searches in the DOM for the corresponding comment button and then clicks it. Afterwards, it searches for the dedicated textarea and write a random sampled comment. Finally, it clicks the button to send the comment.


\section{Sponsored Content Analyses}\label{app:anal}
We report in Table~\ref{tab:sponsoring} the complete overview of audience attracted by our sponsored content. 
In particular, we report overall statistics in term of quantity (e.g., number of likes), and demographic information like gender, age, and location distribution.

\end{document}